\documentclass[11pt,twoside]{article}
\usepackage{asp2004}
\usepackage{psfig}
\usepackage{epsf}
\usepackage{graphics}
\usepackage{lscape}
\newcommand{\timav}{\langle \dot M \rangle}
\newcommand{\be}{\begin{eqnarray}}
\newcommand{\ee}{\end{eqnarray}}

\def\edcomment#1{\iffalse\marginpar{\raggedright\sl#1\/}\else\relax\fi}
\reversemarginpar

\begin{document}
\title{Seismology of Accreting White Dwarfs}
\author{Phil Arras$^1$, Dean Townsley$^2$, Lars Bildsten$^{1,2}$}
\affil{$^1$Kavli Institute for Theoretical Physics, Kohn Hall, University
of California, Santa Barbara, CA 93106 \\
$^2$Department of Physics, Broida Hall, University
of California, Santa Barbara, CA 93106 \\ }

\begin{abstract}
Pulsation modes have recently been observed in a handful of white dwarf
(WD) primaries of cataclysmic variables, allowing an interesting new probe
into the structure of accreting WD's. We briefly discuss the seismology of
these objects, how stellar properties may be inferred from the observed
mode frequencies, and mode driving mechanisms. For one pulsator, GW Lib,
we have shown that a WD mass $M=1.05\ M_\odot$ and accreted envelope
mass $M_{\rm env}=0.4\times 10^{-4}\ M_\odot$ give the best match to the
observed pulsation periods. A first exploration of mode driving favors
$T_{\rm eff} = 14 000 K$ and a massive WD, but more work is necessary.
\end{abstract} \thispagestyle{plain}

\section{Introduction}

Pulsation modes have been observed in isolated white dwarfs for
several decades. Since the observed mode periods are sensitive to the
WD structure, an industry has emerged to estimate parameters such as WD
mass, light element envelope mass, and rotation rate from observed the
mode periods (for a review, see e.g. Bradley 1998). GW
Lib is the first discovered accreting, pulsating white dwarf
\citep{VanZetal00}. It is the shortest orbital period cataclysmic
variable (CV) known at $P_{\rm orb} = 77$ min \citep{Thoretal02}, and is
observed to have dwarf nova outbursts. In the last few years, a handful
of additional accreting pulsators have been found in dwarf nova systems
(see table \ref{tab:periods}) and roughly a dozen more are expected,
raising the possibility of using seismology to determine parameters for
a number of accreting WD in CV's.

Mass determinations for WD in CV's are especially interesting due to possible
implications for progenitor systems of Type Ia supernovae.  Thought to
be accreting WDs near the Chandrasekhar mass \citep{HillNiem00}, these
progenitors are closely related to the CV population, but the precise
nature of this relationship is unknown.  Important clues in this mystery
lie in how the masses of CV primary WDs change over the accreting
lifetime of the binary, but progress is hampered by the difficulty
of measuring CV primary masses \citep{Patt98}.  This mass evolution,
as well as measuring the accreted mass directly with seismology, is also
important for determining how much, if any, of the original WD material
has been ejected into the ISM in classical nova outbursts, contributing
to the ISM metallicity \citep{Gehretal98}.

\begin{table}
\label{tab:periods}
\caption{Known Accreting Pulsating White Dwarfs}
\smallskip
\begin{center}
{\small
\begin{tabular}{ccccc}
\tableline
\noalign{\smallskip}
System & Mode Periods & References & \\
          & [sec] & \\
\noalign{\smallskip}
\tableline
\noalign{\smallskip}
GW Lib & $236$, $377$, $646$ & Van Zyl et.al. (2000,2004) \\
SDSS 1610 & 345, 607 & Warner and Woudt (2004)\\
SDSS 0131 & 330, 600 & '' ''\\
SDSS 2205 & 330, 600  & '' ''\\
SDSS 1556 & 834, 1137, 1558 & Warner and Woudt poster, this meeting \\
SDSS 1556 & 834, 1137, 1558 & '' '' \\
HS 2331 & 60, 300 & Araujo-Betancour et.al (2004) \\
\noalign{\smallskip}
\tableline
\end{tabular}
}
\end{center}
\end{table}

In this proceedings, we first review how seismology can be used to
infer stellar parameters. Next we discuss possible mechanisms for how modes might
be driven in accreting WD. We outline the differences between seismology for the 
accretors and isolated WD.

\section{ Structure and Seismology of Accreting White Dwarfs } 

In this section we first contrast the structure of accreting versus isolated
WD, and then discuss how these differences in structure affect the mode
periods.

Isolated WD are observed to pulsate in narrow ranges of temperature:
$T_{\rm eff} \sim (11-12)\times 10^3{\rm K}$ for hydrogen (DAV) envelopes
\citep{1995ApJ...449..258B}
and $T_{\rm eff} \sim (21-24)\times 10^3{\rm K}$ for helium (DBV).
Sedimentation times near the WD surface are sufficiently short that the
composition is essentially pure H or He for these relatively old WD.
Isolated WD are passively cooling, with the flux being generated by the
hot C/O core. The size of the surface layer of light elements
(H or He) is determined by late stages of nuclear burning and winds,
and is rather poorly constrained by stellar evolution theory
\citep{1990ARA&A..28..139D}.

Envelopes in WD accreting from a companion inherit the componion's
composition. The envelope thickness builds to the ignition mass $ \sim
10^{-4} M_\odot$ at which point a thermonuclear runaway occurs, ejecting
some fraction of the envelope, and the process starts over. Accretion at a
time-averaged rate $\langle \dot{M} \rangle$ generates a ``compressional
heating" luminosity \citep{2004ApJ...600..390T} $L \sim \langle \dot{M}
\rangle k_BT_{\rm core}/m_p$ at the base of the envelope, and determines
the equilibrium core temperature over long ($\sim {\rm Gyr}$) timescales.
The temperature profile differs from an isolated WD since the flux is generated
at the base of the envelope, not uniformly in the core. 

\begin{figure}[!ht]
%\plotone{cavity_integral.eps} 
\plotfiddle{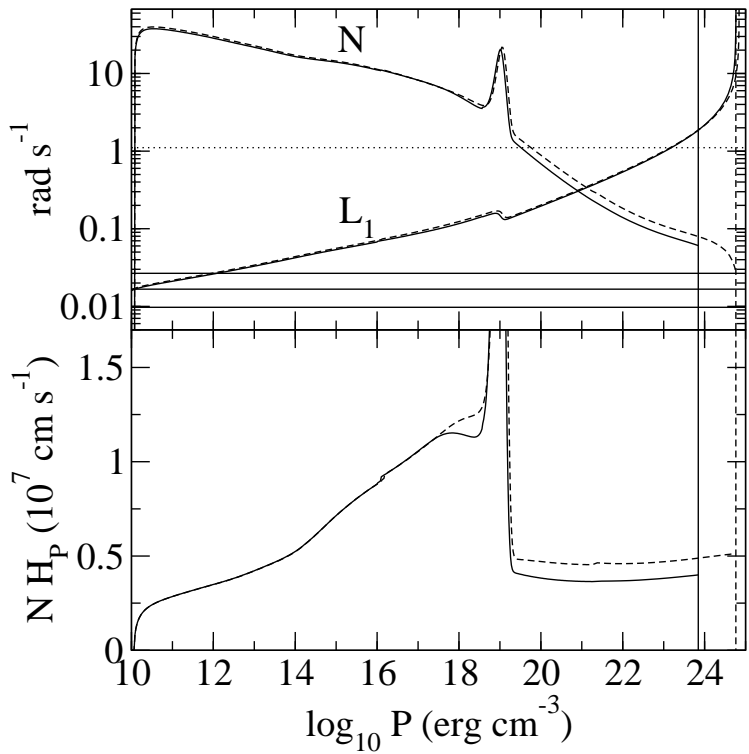}{10.0cm}{0}{80}{80}{-200}{-20}
\caption{ \label{fig:propint}
Comparison of propagation diagram (top panel) and WKB integrand (bottom
panel) for an accreting model (solid lines) and non-accreting model
(dashed line) with the same composition and $T_{\rm eff}$.  The vertical
solid (dashed) line on the right indicates the liquid-solid boundary.
The solid horizontal lines are the three observed frequencies for GW Lib and the
dotted horizontal line is the breakup frequency.  The accreting  
model has $M=1.05M_\odot$, $M_{\rm env}= 0.3M_{\rm ign} = 0.40\times
10^{-4}M_\odot$, $T_{\rm eff}=16070$ K (that implied for this $M$
by the UV spectrum), $\timav = 6.4\times 10^{-11}M_\odot$ yr$^{-1}$,
and $T_{\rm core} = 5.9\times 10^6$ K.  The non-accretor has $T_{\rm
core}=9.3\times10^6$ K.}
\end{figure}

Gravity waves (g-modes) are restored by buoyancy in stably stratified
regions.  The dispersion relation for short wavelength g-modes
is $\omega=Nk_h/k_r$, where $N$ is the Brunt-Vaisalla (buoyancy)
frequency, $k_h^2=l(l+1)/r^2$ is the angular wavenumber for a mode with
angular dependence $Y_{lm}$, and $k_r$ is the vertical wavenumber.
Waves propagate where the vertical wavelength is shorter than the
characteristic lengthscale associated with the background. Hence
waves are reflected from the center and surface, and composition
discontinuities act to create separate resonant cavities, at least for
long wavelength modes. For massive and/or cold WD with solid cores,
g-modes cannot penetrate the core. The propagation diagram for a model
of GW Lib with WD mass $M=1.05M_\odot$ and envelope mass $M_{\rm env} =
0.3M_{\rm ign}=0.40\times10^{-4}M_\odot $ is shown in the upper panel of
Figure \ref{fig:propint}.  The large peak in $N$ at $\log_{10} p\simeq
19$ is due to the change in mean molecular weight in the transition
layer from the solar composition accreted envelope to the C/O core.
In the roughly constant flux envelope, $N^2 \simeq g/z$ where $z$ is the
depth into the star from the surface. In the degenerate core, $N^2 \simeq
(g/H_p)(k_bT/E_F)$ drops rapidly toward the center, where $E_F$ is the
electron Fermi energy and $H_p = p/\rho g$ is the pressure scale height.

The contribution to the eigenfrequency from a particular part of
the star can be estimated from the WKB phase integral $\int dr k_r =
[l(l+1)]^{1/2} \omega^{-1} \int dr N(r)/r \approx n\pi$, hence $\omega
\propto \int dr N(r)/r \propto \int d\ln p H_p N$, where we have taken
the radius to be approximately constant in the last expression. The
integrand is shown in the bottom panel of Figure \ref{fig:propint},
representing the number of nodes per decade in pressure.  Shown are two
WD models: an accreting model with $M=1.05M_\odot$ and $M_{\rm env} =
0.3M_{\rm ign}=0.40\times 10^{-4}M_\odot$ used in our mode analysis
(solid line), and a non-accreting model with the same $T_{\rm eff}$
(dashed line) and composition. Not shown in Figure \ref{fig:propint},
but perhaps the largest effect, is the difference due to the envelope
mean molecular weight, $\mu$, from the case of a pure H or He envelope
to one of solar compositon.  Such a change is reflected in the periods as
$P_n\propto\mu^{0.5}$, since the envelope is well-approximated by an $n=4$
polytrope.  The non-accreting model also has a higher $T_{\rm core}$ than
the accreting model by about $50\%$, leading to two important effects:
(1) the WKB integrand has a higher value in the core for the non-accreting
model ($\omega \propto T_{\rm core}^{1/2}$), and (2) the solid core is
smaller, pushing the inner boundary condition deeper into the star. Both
effects directly influence the observed periods, and period spacings,
hence {\it it is essential to use a WD model including the effects of
compressional heating and nuclear burning}, rather than a passively
cooling WD model.

To get an analytic idea for how stellar parameters can be estimated from
mode frequencies, consider first the nondegenerate envelope. In this
region, $N^2 \sim g/z$ giving a frequency $\omega \sim [l(l+1)]^{1/2}
n^{-1} (g/H)^{1/2} (H/R)$ for a mode with quantum numbers $l$ and
$n$ trapped in an envelope with base scale height $H$ and stellar
radius $R$. For a mode trapped in the degenerate core, $\omega \sim
[l(l+1)]^{1/2} n^{-1} (g/H)^{1/2} (k_BT_{\rm core}/E_F)^{1/2}$. Given
observed mode periods, and assumptions about the $l$ quantum number, one
can attempt to identify the $n$ quantum number of the modes, and infer
stellar parameters such as mass and envelope mass using these expressions.

Townsley, Arras, and Bildsten (2004) have used the model of accreting WD
thermal structure from Townsley and Bildsten (2004) in the first study
of seismology of accreting WD. They computed oscillation mode periods
using an adiabatic code, and identified the WD model most consistent
with the three observed mode periods (assumed to be dipole) for GW Lib,
the best observed accreting pulsating WD. The best fit was a WD mass
$M=1.05M_\odot$ and accreted envelope mass $M_{\rm env}= 0.3M_{\rm ign}
= 0.40\times 10^{-4}M_\odot$. Such a large inferred WD mass has consequences
for mode excitation, discussed in the next section.

\section{ Excitation of G-Modes in Accreting White Dwarfs }

We are motivated to understand how mode excitation might be different
in accreting WD for two reasons. First, Szkody et.al. (2003) have
determined an effective temperature for GW Lib of $T_{\rm eff} = (14-17)
\times 10^3\ {\rm K}$, depending on how the fit was done. These estimates
are well above the standard DAV instability strip $T_{\rm eff} \approx
(11-12)\ \times 10^3\ {\rm K}$, implying that convective driving, and
also the $\kappa$-mechanism, cannot drive the pulsations for a fiducial
$M=0.6M_\odot$ pure hydrogen envelope. Secondly, the WD mass may have
increased after ${\rm Gyr}$'s of accretion, motivating study of WD
heavier than the canonical $M=0.6M_\odot$.  
%Third, all the accreting
%pulsators found so far have near solar composition companions. However,
%below the period gap the donor star is expected to be hydrogen-depleted,
%implying the mode driving calculations need to be redone for a range
%of composition.

%\begin{figure}[!ht]
%\plotone{tthbcz_XH.ps}
%\plotfiddle{tthbcz_XH.ps}{10.0cm}{0}{50}{50}{-150}{-75}
%\caption{ \label{fig:tthbczXH}
%Thermal time at base of convection zone vs. $T_{\rm eff}$ for different H
%to He ratios, but fixed gravity $g=10^8\ {\rm cm\ s^{-2}}$ and metallicity
%$Z=0.02$. Note there can be two convections zones associated with second
%He ionization (upper right hand corner), or first He ionization (lower
%right hand corner) and/or H ionization (nearly vertical lines starting
%at the left). The H mass fraction for each line is X=0.98 (solid black),
%0.90 (solid red), 0.80 (solid green), 0.70 (solid blue), 0.60 (solid
%cyan), 0.50 (solid magenta), 0.40 (solid yellow), 0.30 (dashed black),
%0.20 (dashed red), 0.10 (dashed green), 0.00 (dashed blue). } \end{figure}

A large literature exists on excitation of g-modes in isolated
WD. Initially investigators expected the ``$\kappa$-mechanism", associated
with opacity variations in an ionization zone, was the driving mechanism
\citep{1981AA....97...16D}. Later, with penetrating insight, Brickhill
realized that the response of surface convection zones to a g-mode
pulsation naturally acts to pump more energy into the mode, hence the name
``convective driving" \citep{1983MNRAS.204..537B,1991MNRAS.251..673B}.
Wu and Goldreich have confirmed Brickhill's result, and extended its
consequences, notably in the area of nonlinear saturation mechanisms (see
the series of papers starting with Goldreich and Wu 1999).  The criterion
for a mode to be unstable is roughly $ \omega \tau_{\rm th} \geq 1$,
where $\omega=2\pi/P$ is the mode frequency, $P$ is the period, and
$\tau_{\rm th}$ is the thermal time at the base of the ionization zone,
which coincides with the base of the convection zone. The ``blue edge"
of the instability strip in effective temperature occurs when the
shortest period g-mode is unstable by this criterion; g-modes should
be observed for temperatures below the blue edge. For DAV and DBV's,
observationally there is a  ``red edge" below which pulsations are
not observed. The reason for this cutoff is not as well understood,
but may have to do with the luminosity perturbation becoming small,
even if the mode amplitude is large (Brickhill 1983).

\begin{figure}[!ht]
%\plotone{tthbcz_g.ps}
\plotfiddle{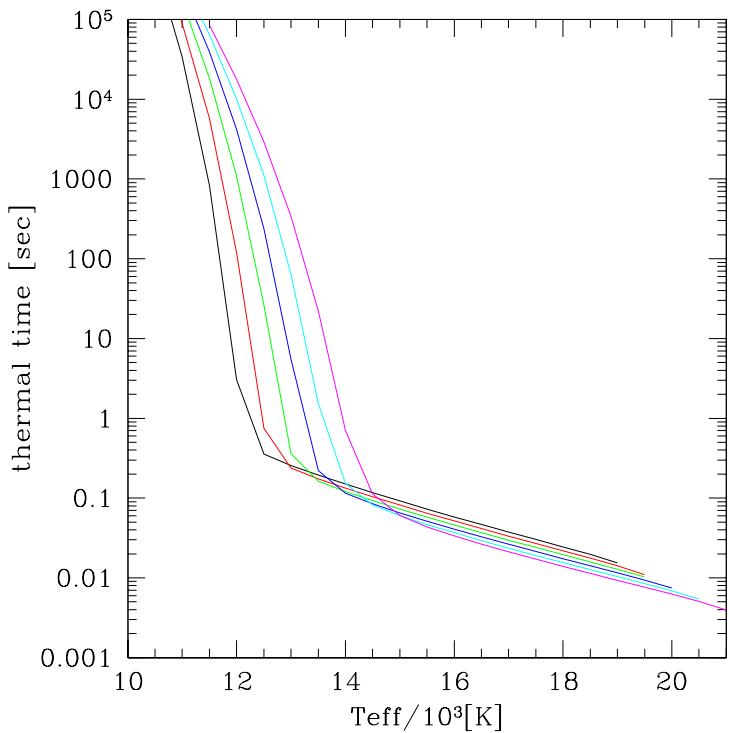}{10.0cm}{0}{50}{50}{-150}{-75}
\caption{ \label{fig:tthbczg}
Thermal time at base of convection zone vs. $T_{\rm eff}$ for different
gravity, but fixed (solar) composition $X=0.7$ and $Z=0.02$.  The gravity
in cgs units for each line is, from left to right, $g=10^{8.0}$ (black),
$10^{8.2}$ (red), $10^{8.4}$ (green), $10^{8.6}$ (blue), $10^{8.8}$
(cyan), $10^{9.0}$ (magenta).}
 \end{figure}

What is the instability strip for accreting WD?  Here we investigate
the thermal time at the base of the convection zones as a function
of %hydrogen to helium ratio (fig.\ref{fig:tthbczXH}) and surface
gravity (fig.\ref{fig:tthbczg}). We use the OPAL opacity and equation
of state \citep{1996ApJ...464..943I} to construct solar composition,
plane parallel, constant gravity and flux envelopes, using standard
mixing length \citep{1990sse..book.....K} in convective regions. In
fig.\ref{fig:tthbczg}, gravity is varied from $g=10^8{\rm cm\ s^{-2}}$
($M \approx 0.6 M_\odot$) to $g=10^9{\rm cm\ s^{-2}}$ ($M \approx 1.2
M_\odot$). We define a fiducial blue edge to be at $t_{\rm th}=100\
{\rm sec}$. This blue edge moves to the right by $\sim 2000\ {\rm K}$
over a factor of 10 change in gravity.

%In fig.\ref{fig:tthbczXH}, starting at the left hand side for the
%hydrogen convection zone, we see that the blue edge ($t_{\rm th}$ greater
%than, say, $100{\rm sec}$ as a fiducial case) moves to higher $T_{\rm
%eff}$ as the hydrogen content decreases. The blue edge increases by
%$\sim 3000\ {\rm K}$ as the hydrogen content drops from solar to half
%solar. As the hydrogen content drops to zero we recover the blue edge
%for the DBV's, near $T_{\rm eff} \approx 25000{\rm K}$.

We confirm the suggestion by Szkody et.al. (2003) that GW Lib may pulsate
because it is a massive WD. If GW Lib has a high surface gravity such that
the blue edge moves up to $T_{\rm eff} \approx 14000{\rm K}$, it may overlap
with the lower value stated for the observed $T_{\rm eff}$. A large WD mass
$M \approx 1.1 M_\odot$ was also preferred in the parameter estimation of 
Townsley et.al. (2004) using seismology.

Additional factors not discussed here may influence seismology and g-mode
driving in accreting WD. The structure of composition transition layers
is different for accretors due to the downward accretion flow, and the
composition of layers overlying the core is uncertain due to classical
novae. Hydrogen depleted donor stars are expected for binaries with
orbital periods $\la 1\ {\rm hr}$, increasing the blue edge somewhere
in between that of DAV's and DBV's. Metals may act as an additional {\it
damping} mechanism (cool WD are on the wrong side of the ``metal bump").
Sedimentation in between dwarf novae outbursts may occur in the accreted
near solar composition layer, even down to the driving region at $p \sim
10^{10}\ {\rm erg\ cm^{-3}}$, creating helium ``puddles" sandwiched in
between hydrogen and solar composition layers.  Seismology for WD spun up
even to modest rotation rates will be quite different than the nonrotating
case, changing the expected mode periods qualitatively. Finally, new
instability strips may occur outside the traditional DAV and DBV regions
due to changes in mode properties arising from rapid rotation.

\acknowledgements We thank Tony Piro, Phil Chang, Boris G\"ansicke,
and Steve Howell for useful conversations and comments. This work was
supported by the National Science Foundation under grants PHY99-07949
and AST02-05956. Phil Arras is a NSF AAPF fellow.

\end{document}